\documentclass[lettersize,journal]{IEEEtran}
\usepackage{amsmath,amsfonts}
\usepackage{algorithmic}
\usepackage{algorithm}
\usepackage{array}
\usepackage[caption=false,font=normalsize,labelfont=sf,textfont=sf]{subfig}
\usepackage{textcomp}
\usepackage{stfloats}
\usepackage{url}
\usepackage{verbatim}
\usepackage{graphicx}
\usepackage{cite}
\usepackage{balance}
\usepackage{array}
\usepackage{pifont}
\usepackage[table]{xcolor}
\begin{document}

\title{Quantum Internet in the Sky: Vision, Challenges, Solutions, and Future Directions}

\author{Phuc V. Trinh,~\IEEEmembership{Member,~IEEE,} and Shinya Sugiura,~\IEEEmembership{Senior Member,~IEEE}
% <-this % stops a space
%\vspace*{-7mm}
\thanks{Preprint (Accepted Version). DOI: 10.1109/MCOM.003.2300835.
Copyright $\copyright$ 2024 IEEE. Personal use of this material is permitted. However, permission to use this material for any other purposes must be obtained from the IEEE by sending a request to pubs-permissions@ieee.org.}
\thanks{The authors are with the Institute of Industrial Science, The University of Tokyo, Tokyo 153-8505, Japan. (\textit{Corresponding author: Shinya Sugiura}.)} %(e-mail: \{trinh, sugiura\}@iis.u-tokyo.ac.jp). }% <-this % stops a space
\thanks{This work was supported in part by the JSPS KAKENHI (23H00470), in part by the JST FOREST (JPMJFR2127), and in part by the National Institute of Information and Communications Technology, Japan.} % (NICT), Japan.}
%\thanks{Manuscript received April 19, 2021; revised August 16, 2021.}
}

% The paper headers
\markboth{Preprint (Accepted Version) for publication in IEEE Communications Magazine (DOI: 10.1109/MCOM.003.2300835)}%IEEE Communications Magazine}%
{}

%\IEEEpubid{0000--0000/00\$00.00~\copyright~2021 IEEE}
% Remember, if you use this you must call \IEEEpubidadjcol in the second
% column for its text to clear the IEEEpubid mark.

\maketitle

\begin{abstract}
This article envisions the concept of a ``Quantum Internet in the Sky", aiming to establish ubiquitous quantum communication links among distant nodes via free-space optical channels. Our key focus is on deploying quantum communication terminals on non-terrestrial platforms, specifically unmanned aerial vehicles and satellites, at various altitudes. By highlighting the unique characteristics of these platforms compared to terrestrial counterparts, we address inherent challenges and discuss potential solutions through meticulous system designs and analyses of typical non-terrestrial quantum communication scenarios. Finally, we illuminate the path forward by proposing essential future directions that underscore the integration of high-dimensional multipartite quantum communications with sensing, computing, and intelligence for multiple users en route to realizing a fully operational Quantum Internet.
\end{abstract}


\section{Introduction}
\IEEEPARstart{R}{ecent} advancements in quantum computers play a crucial role in expanding the limits of computational capabilities and transformative applications. Quantum computers use quantum bits (qubits), relying on the principles of quantum superposition to exist in multiple quantum states that represent classical bits 0 and 1 simultaneously, allowing for computations surpassing the efficiency of classical computers. The development began with theoretical foundations laid in the 1980s by Feynman and Deutsch, who envisioned quantum computers as tools for simulating quantum systems \cite{feynman1982} and formulated the principles of quantum computation \cite{deutsch1985}, respectively. In 1994, Shor invented a pioneering quantum algorithm for factoring large numbers exponentially faster than classical counterparts \cite{shor1994}, which could be used to break public-key cryptography schemes. A groundbreaking milestone of \textit{quantum supremacy} was then achieved by Google in 2019 \cite{arute2019}, practically demonstrating the capacity of quantum computers to outperform classical ones in specific tasks.

As quantum computers evolve, their integration with communication networks becomes increasingly feasible, driving the transition from quantum computing breakthroughs to the realization of a revolutionary ``Quantum Internet". It is expected to work in synergy with the current classical Internet to enable an unparalleled level of security, distributed computing, and information processing. Among these key elements of the Quantum Internet, security stands out prominently, exemplified by the application of quantum key distribution (QKD). By using quantum states to distribute cryptographic keys between parties, QKD ensures that any attempt by eavesdroppers to intercept and copy the transmitted quantum states will inevitably disturb the states, thus revealing their presence. In distributed computing, simultaneous manipulations of qubits across distributed devices can be realized by utilizing entanglement swapping, which allows for the creation of entangled ``virtual" connections between nodes lacking direct physical links. This offers increased processing power through parallelization, improved fault tolerance, and optimized resource allocations. In information processing, the transmission of quantum information over long distances can be facilitated by quantum teleportation, enabling the instantaneous transfer of quantum states without physical particle movements.

Recent trends witness a pivotal shift in quantum network architecture, transitioning from terrestrial infrastructures to seamless integration with non-terrestrial platforms encompassing unmanned aerial vehicles (UAVs) and satellites. This aligns synergistically with the evolution of non-terrestrial networks (NTNs) within the framework of the sixth-generation (6G) ecosystem. To take one step closer to reality, Chinese scientists have established the first integrated quantum communication network, combining over 700 optical fiber links in a trusted-relay structure on the ground and two satellite-to-ground QKD links over a total distance of 4,600 km \cite{chen2021}. Inspired by this success, integrating quantum terrestrial networks with the burgeoning applications of NTNs brings forth the potential for ubiquitous and globally interconnected quantum infrastructures. This transformation will not only reshape the dynamics of worldwide communications but also exert a profound influence on various sectors, including industry, finance, administration, and operational and fundamental sciences.

This article aims to deliver a visionary concept of a \textit{Quantum Internet in the Sky} tailored for quantum NTNs. Our central contribution involves addressing the inherent challenges posed by non-terrestrial platforms, which have a significant impact on the key elements of the Quantum Internet, including security, computing, and information processing. Potential solutions are discussed via exemplary system designs and numerical analyses. Finally, we identify future directions.
%========================================%
\section{Vision of Quantum Internet in the Sky}
\begin{figure*}[t]
\centering
\includegraphics[scale=0.45]{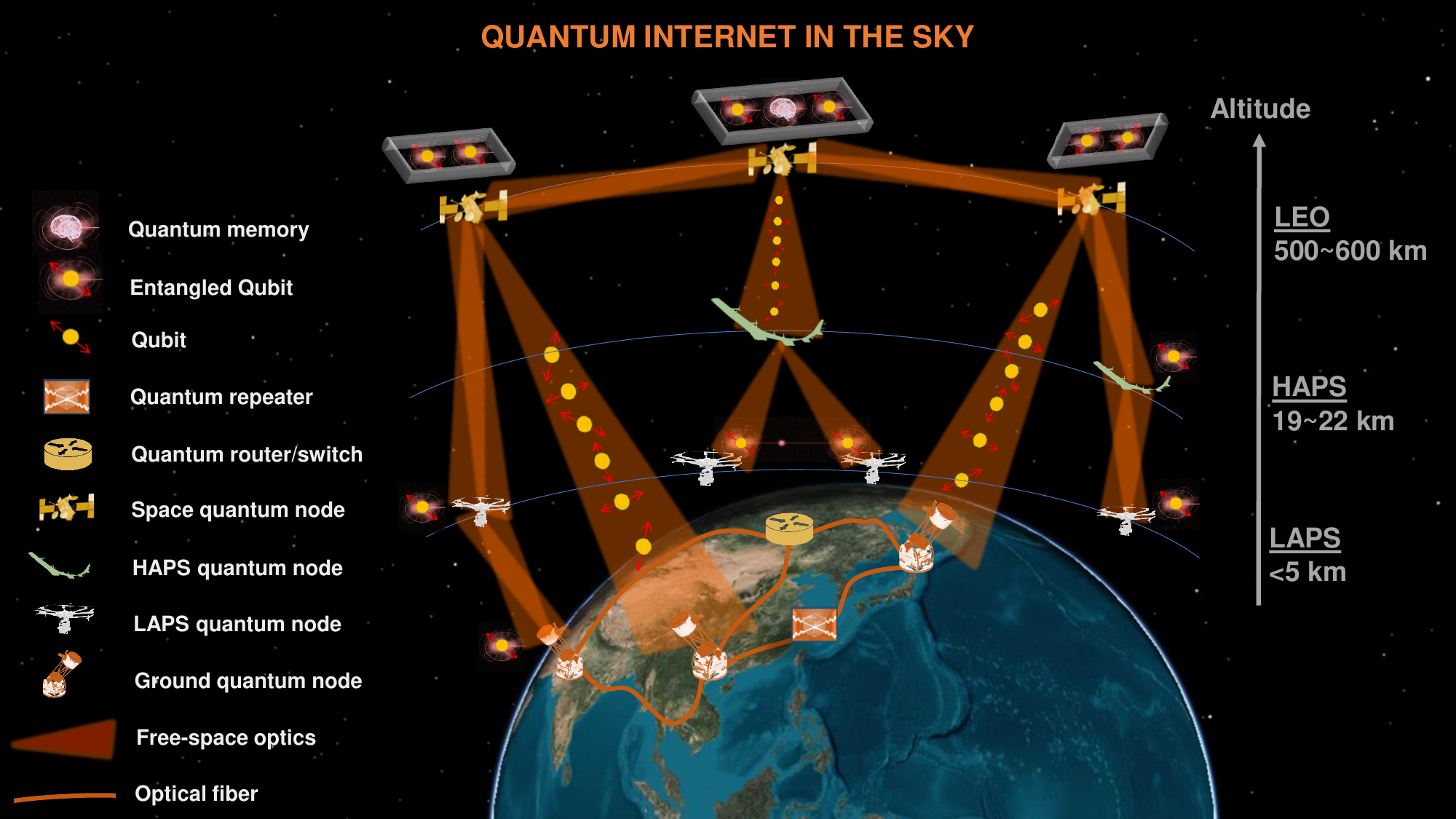}
\caption{Overview concept of the ``Quantum Internet in the Sky".}
\label{fig_1}
\end{figure*}
In the Quantum Internet, photonic qubits, employing photons as carriers of quantum information, stand out for their compatibility with existing optical fiber and free-space optical (FSO) networks. Particularly, FSO systems employ laser beams to transmit data through the air or space, establishing direct line-of-sight (LoS) connections without physical cables. This inherent feature makes FSO notably appealing for NTNs, serving as a cost-effective and easily deployable means of connecting distant nodes.

Our vision of \textit{Quantum Internet in the Sky} emphasizes the transmission of photonic qubits through FSO systems integrated on non-terrestrial platforms at various altitudes, including low-altitude platform stations (LAPS), high-altitude platform stations (HAPS), and low-Earth orbit (LEO) satellites, forming an intricate three-dimensional (3D) mesh network as illustrated in Fig. \ref{fig_1}. In this conceptual framework, entanglement resources are harnessed to convey quantum information across multi-layer non-terrestrial platforms and distributed into the ground networks. Initially, entanglement links are dynamically established within each layer and across layers of LEO satellites, HAPS, and LAPS, then extended to ground segments. In response to user requests, end-to-end entanglement is formed for any desired nodes within the integrated non-terrestrial and terrestrial quantum network through entanglement swapping operations. Finally, end users on ground or non-terrestrial platforms leverage this entanglement resource to fulfill their quantum communication requirements. Once end-to-end entanglement is established, quantum information does not hop through intermediate nodes but directly from one end node to another, regardless of their distant locations. The operations specific to each non-terrestrial layer are outlined as follows.

A layer composed of numerous satellites operating in LEO, commonly referred to as LEO satellite constellations and exemplified by initiatives like Starlink, could enable global quantum communications by leveraging its extensive coverage and low-latency characteristics. Additionally, quantum signal experiences significantly reduced losses over free-space channels, enabling direct links spanning thousands of kilometers rather than some hundred kilometers achievable over optical fiber. Each LEO satellite fosters secure communication links between distant HAPS/LAPS or ground stations by serving as a trusted or untrusted relay, depending on the selected QKD protocols. For communications over intercontinental distances, multiple satellites within the constellation can be assigned as quantum repeaters, relying on entanglement swapping to extend the communication range. Entanglement swapping is a quantum phenomenon where the entanglement of two qubits is transferred to a pair of qubits that have never directly interacted, accomplished through a joint Bell-state measurement and correlation of intermediate qubits. It may also require that these satellites are equipped with quantum memories to temporarily store entangled qubits before being retransmitted. In order to enable routing and switching within the constellation, some designated satellites can act as quantum routers and switches, which allow for routing or switching entanglement without copying information. Quantum routers utilize entanglement-assisted routing, while quantum switches dynamically adapt qubit paths through quantum gates. For the sake of mitigating quantum decoherence effects and noises during qubit transmissions, quantum error correction algorithms should be implemented. Further entanglement purifications can also improve the fidelity of entangled states.
A layer consisting of aerial platforms that typically operate at altitudes ranging from 19 km to 22 km in the stratosphere is referred to as HAPS. These platforms are designed to remain quasi-stationed in the air for extended periods by harnessing solar power, offering various applications across communications, surveillance, and remote sensing. When equipped with quantum communication terminals, HAPS can help to extend the reach and coverage of quantum networks under challenging terrains. The strategic positioning of HAPS allows for longer link durations with multiple satellites and being unaffected by atmospheric interference, while serving as trusted/untrusted relay nodes to lower-altitude platforms such as LAPS and ground stations. Similar to inter-satellite quantum links, multiple HAPSs can function as quantum repeaters using entanglement swapping to achieve collectively broader transmission coverages.
A layer containing UAVs, such as fixed-wing or rotary-wing drones, flying or hovering at altitudes below 5 km, is categorized as LAPS. These platforms provide a dynamic and agile infrastructure for quantum communications, suitable for immediate deployment in disaster zones and complex urban environments. Unlike HAPS with extended flight durations, drones can only serve as temporary relay nodes, enabling quantum communication links in specific on-demand regions and environments. This limitation stems from the trade-off between the payload-carrying capacity and the operational time, constrained by the energy stored in drone batteries. The interconnectivity within a swarm of drones is anticipated to dynamically adjust to real-time weather conditions, ensuring continuous communication coverage. For example, depleted drones can be seamlessly substituted with fully charged ones to sustain connectivity. Additionally, merging multiple swarms can broaden communication ranges and bolster resilience against adverse weather conditions. This capability proves invaluable in addressing the last-mile quantum key exchange challenge for inner-city or field networks.
%========================================%
\section{Key Challenges \& Potential Solutions}
\begin{table*}[t]
\caption{Key challenges for typical scenarios of non-terrestrial quantum communications.}
\label{Table_1}
\centering
\begin{tabular}{|l|c|c|c|c|} \hline
\textbf{Challenges} &\textbf{LEO-to-Ground} &\textbf{LEO-to-HAPS} & \textbf{LEO-to-LAPS} & \textbf{HAPS-to-LAPS} \\ \hline \hline
Polarization reference-frame misalignments and phase delay \cite{carrasco2016} &\checkmark &\checkmark &\checkmark &\checkmark \\ \hline
Clock data recovery and timing-offset identification \cite{takenaka2017} &\checkmark &\checkmark &\checkmark &\checkmark \\ \hline
Finite-key effects in QKD systems \cite{sidhu2022}&\checkmark &\checkmark &\checkmark &\checkmark \\ \hline
Pointing errors and geometrical losses \cite{trinh2022}&\checkmark &\checkmark &\checkmark &\checkmark \\ \hline
Background noise \cite{khatri2021}&\checkmark &\checkmark &\checkmark &\checkmark \\ \hline
Atmospheric turbulence \cite{trinh2022,vasylyev2018} &\checkmark &\cellcolor[HTML]{C0C0C0} &\checkmark &\checkmark \\ \hline
LEO satellite slew rates \cite{trinh2022} &\checkmark &\checkmark &\checkmark &\cellcolor[HTML]{C0C0C0} \\ \hline
Cloud blockage \cite{birch2023} &\checkmark &\cellcolor[HTML]{C0C0C0} &\checkmark &\checkmark \\ \hline
Miniaturized terminal \cite{carrasco2022} &\cellcolor[HTML]{C0C0C0} &\checkmark &\checkmark &\checkmark \\ \hline
\end{tabular}
\end{table*}
In non-terrestrial quantum communications, downlink refers to communications from platforms at higher altitudes to lower altitudes, and uplink represents the reverse direction. Table \ref{Table_1} concisely summarizes the challenges across typical inter-platform quantum communication scenarios, extracted from Fig. \ref{fig_1}. It is essential to recognize that all non-terrestrial platforms have the capability to establish quantum links with ground stations. However, we narrow our focus to the LEO-to-ground link, notable for its extended path to the ground and shared characteristics with HAPS/LAPS-to-ground links. Thereafter, we explore representative challenges and solutions for each scenario.
\subsection{LEO Satellite-to-LAPS Quantum Links}
\subsubsection{Uplink/Downlink Challenges}
\label{subsect:LEO_LAPS_challenges}
In this scenario, both uplink and downlink suffer from atmospheric turbulence-induced quantum decoherence \cite{vasylyev2018}, pointing errors (PE) and geometrical losses over long distances \cite{trinh2022}, and background noise \cite{khatri2021}. Under comparable atmospheric conditions, the uplink suffers more severe beam spreading and beam wandering. %, leading to intolerable losses.
Especially, beam wandering yields random displacements of the instantaneous beam center, which can cause the uplink beam to miss the target satellite. Due to relative motions between satellites and LAPS, Doppler shift becomes a serious concern as it causes frequency mismatches affecting clock data recovery and timing-offset identification \cite{takenaka2017}. Regarding QKD operations, the battery-constrained flying duration of LAPS and the restricted time window during an LEO satellite overpass result in a limited number of quantum signal transmissions, leading to statistical fluctuations known as finite-key effects \cite{sidhu2022}. These fluctuations can deviate from the expected values predicted by idealized models, deteriorating the estimation of key generation parameters. Last but not least, the satellite and LAPS motions make the polarization reference frame change with time, inducing basis misalignments between two communication ends. Moreover, the varying incident angle inside receiving optics introduces frequency-dependent phase delays between orthogonal polarization components, possibly turning the linear polarization into elliptical \cite{carrasco2016}.
\subsubsection{Uplink/Downlink Solutions}
\label{subsect:LEO_LAPS_solutions}
Due to the stringent constraints of size, weight, and power consumption (SWaP) on micro-satellites and LAPS, miniaturized quantum communication terminals tailored for such small platforms are required, while incorporating both gimbal-based coarse-pointing and fine-pointing/tracking subsystems \cite{carrasco2022}. Coarse-pointing orients the gimbal roughly while fine-pointing/tracking employs a closed-loop system with fast-steering mirrors and position-sensing detectors for precise beam alignments between the two dynamic platforms. After the laser beam is stabilized, a very narrow beam-divergence angle is desirable to minimize the geometrical losses due to limited receiving aperture sizes, typically $10\!\sim\!15$ cm, on both micro-satellite and LAPS. However, the divergence angle for the uplink needs to be wide enough to cope with beam-wander movements due to atmospheric effects, and the point-ahead angle from LAPS must be accurately calculated to ensure that the moving satellite can be reached by the uplink. To receive the correct quantum states, it is important to track polarization reference frames between satellites and LAPS by placing a motorized half-wave plate at the receiver to rotate the polarization in response to satellite and LAPS orientations \cite{takenaka2017}. Additionally, clock data recovery and timing-offset synchronization can be achieved by directly extracting information from the received quantum states during the post-processing steps \cite{takenaka2017}. Finally, to optimize LAPS for on-demand quantum links under SWaP constraints, a practical approach is to directly couple the received free-space beams onto single-photon avalanche detectors (SPADs) instead of sophisticated fiber-coupling ones to avoid additional coupling losses.
\subsection{LEO Satellite-to-HAPS Quantum Links}
\subsubsection{Uplink/Downlink Challenges}
This scenario promises efficient quantum connections due to the absence of atmosphere in the stratosphere, thus avoiding the severe effects of atmospheric turbulence \cite{vasylyev2018} and cloud blockage \cite{birch2023}. However, HAPS is also imposed by relatively tight SWaP constraints, hindering the deployment of costly payloads with large telescope apertures and bulky signal-processing electronics. This limitation dictates the use of compact telescopes, potentially with diameters of up to 35 cm, within a total payload weighing less than 70 kg. On the satellite side, the telescope size is much more restricted, particularly with the proliferation of future nano-satellites like CubeSats, reducing the telescope size to less than 10 cm. As a result, achieving acceptable geometrical losses between the widened beam footprint and the receiving telescope aperture is crucial for optimal system designs \cite{trinh2022}. As for QKD, finite-key effects \cite{sidhu2022} are unavoidable due to the brief communication time with LEO satellites despite the extended operational periods of HAPS. Besides, this scenario presents analogous challenges in link acquisitions, tracking and pointing (ATP) \cite{carrasco2022}, reference-frame alignments \cite{carrasco2016} and timing synchronization \cite{takenaka2017} as observed in the LEO satellite-to-LAPS scenario for both uplink and downlink, owing to the quasi-stationary nature of both HAPS and LAPS.
\subsubsection{Uplink/Downlink Solutions}
\label{subsect:LEO_HAPS_solutions}
Similar to Section \ref{subsect:LEO_LAPS_solutions}, the recommended solutions for addressing challenges in LAPS are equally applicable to HAPS. Here, to give further insights into the QKD performance over a LEO satellite-to-HAPS downlink, we will present a numerical investigation focusing on the secret-key length (SKL) of a decoy-state efficient Bennett-Brassard 1984 (BB84) QKD protocol considering finite-key effects, following approaches presented in \cite{sidhu2022,trinh2022}. Particularly, we consider a CubeSat in LEO hosting a transmitting Cassegrain telescope with a 9-cm aperture that produces a full-angle divergence of 33 $\mu$rad \cite{trinh2022} and a HAPS carrying a 35-cm Rx telescope. The CubeSat is assumed to be in a Sun-synchronous orbit at an altitude of 535 km, while the HAPS is deployed at a 20 km altitude in Tokyo, Japan. The link and system parameters are similar to \cite[Table 2]{trinh2022}, except that there is no atmospheric loss, and the optical wavelength and QKD source rate are 810 nm and 200 MHz, respectively.

\begin{figure}[t]
\centering
\includegraphics[scale=0.58]{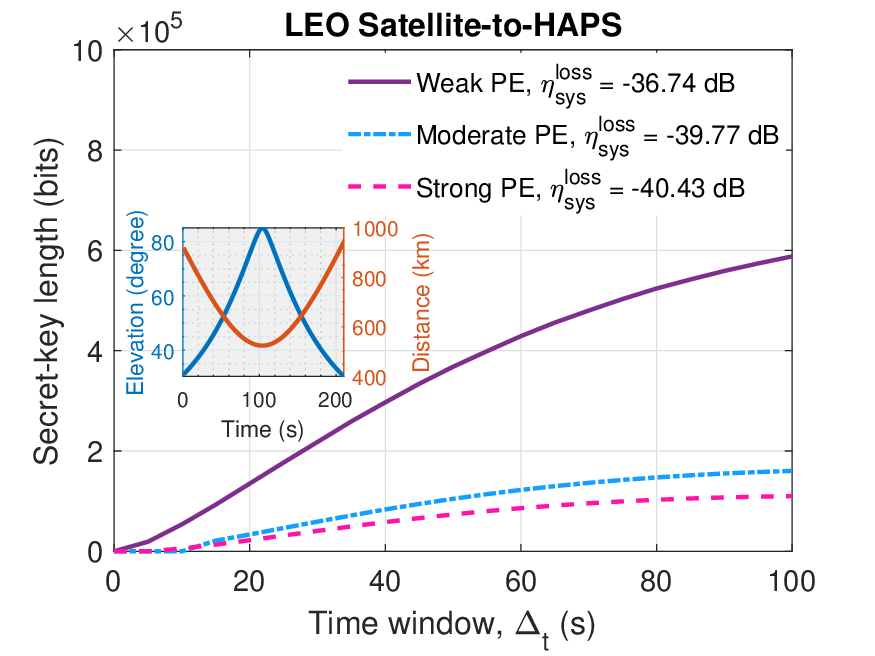}
\caption{Secret-key length of a decoy-state efficient BB84 QKD protocol with finite-key effects for a LEO satellite-to-HAPS downlink at 810 nm.}
\label{fig_2}
\end{figure}
Fig.~\ref{fig_2} shows the SKL versus the LEO-to-HAPS communication time half-window $\Delta_{t}$ of 100 s, where the SKL extractable from received quantum data within the full-window duration of $2\Delta_{t}$ is optimized over QKD protocol parameters for each value of $\Delta_{t}$. Various PE levels defined in \cite{trinh2022} are plotted to highlight the impact of unstable platforms on the link alignments that eventually affect the total system loss denoted as $\eta_{\text{sys}}^{\text{loss}}$. The simulated elevation and communication distance over the satellite pass are presented in the inset in Fig. \ref{fig_2}. It is seen that moderate and strong PE levels exert significant losses on the system, restricting the final SKLs to below 0.2 Mbits. Meanwhile, a weak PE level permits a higher accumulated SKL, reaching approximately 0.6 Mbits at the end of the satellite pass. This highlights the importance of a precise pointing/tracking subsystem in achieving a longer SKL, exemplified by a beam-jitter standard deviation of 3.3 $\mu$rad in the weak-PE case. A longer SKL facilitates secure encryption of larger data volumes, adhering to the one-time pad principle where the key length must equal or surpass the data size for maximum security.
\subsection{HAPS-to-LAPS Quantum Links}
\subsubsection{Uplink/Downlink Challenges}
In interconnecting SWaP-constrained platforms like HAPS and LAPS, ensuring a robust link is crucial for a high quantum fidelity amid the diverse day and night conditions and varying background noise levels. Unlike fast-moving LEO satellites, HAPS and LAPS are quasi-stationary or cruising at low speeds of $10\!\sim\!30$ m/s, making the link ATP easier in both uplink and downlink. Although this scenario still endures atmospheric effects in the first 20 km of the atmosphere layer, the impact is modest as atmospheric turbulence is less severe at high altitudes. The primary challenge, however, lies in the limitations posed by the use of small telescopes on both HAPS and LAPS. This is due to the fact that smaller telescopes have a wider field of view (FoV), thus leaking more background noise into the received quantum signals. Simultaneously, geometrical losses are inevitable due to the small size of the telescope aperture compared to the broadened beam footprint. This considerably diminishes quantum fidelity, particularly during daytime operations with increased spectral irradiance.
\subsubsection{Uplink/Downlink Solutions}
While previous solutions offered in Sections \ref{subsect:LEO_LAPS_solutions} and \ref{subsect:LEO_HAPS_solutions} are applicable in dealing with the challenges in HAPS and LAPS platforms, a more thorough investigation into the impact of varying spectral irradiance levels and geometrical losses is necessary to quantify their effects on quantum fidelity, which is an important indicator to assess the similarity between two quantum states. We now explore a case study where a HAPS at 20 km altitude generates an entangled photon pair and distributes individual photons to two distant LAPSs at 1 km altitude in the presence of background photons stemming from solar irradiance and diffuse sky radiation. The receiving telescope at both LAPSs maintains a zenith angle of $40^{\circ}$ with respect to HAPS, resulting in a communication distance of 24.8 km from each LAPS to the HAPS. The entanglement transmission fidelity is subsequently assessed as a function of spectral irradiance following the methodology in \cite{khatri2021}, with additional considerations of various PE levels as defined in \cite{trinh2022}.

\begin{figure}[t]
\centering
\includegraphics[scale=0.58]{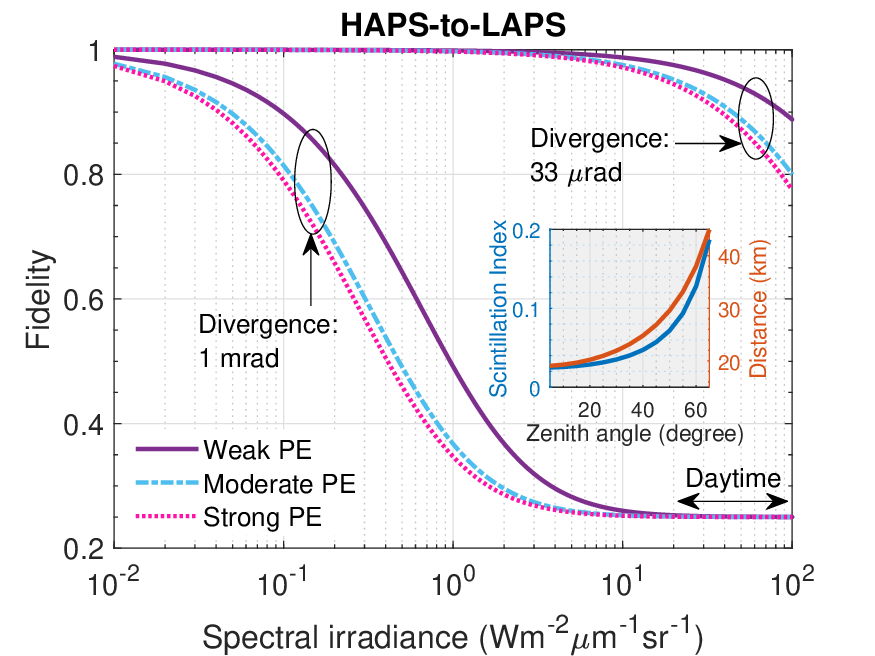}
\caption{Fidelity of a HAPS-to-LAPS downlink entanglement transmission as a function of spectral irradiance at 810 nm under various PE levels.}
\label{fig_3}
\end{figure}
The fidelity results are depicted in Fig. \ref{fig_3}, where the inset illustrates atmospheric turbulence severity, measured by the scintillation index (SI) parameter, versus zenith angles and communication distances between HAPS and LAPS. It is observed that SI is much less than unity in all practical zenith angles and communication distances, suggesting a minimal impact of turbulence. Qualitatively, two observations can be drawn from Fig. \ref{fig_3}: (i) the effects of PE severities under large geometrical losses are dominated by that of the high spectral irradiance levels; (ii) employing a narrow full-angle transmitting divergence, such as 33 $\mu$rad, proves effective in achieving significantly low geometrical losses compared to a 1-mrad divergence. This ensures a high fidelity, approximately exceeding 80\%, even in daytime conditions. However, utilizing a narrow beam may increase link outage probability in severe weather and operating conditions. Therefore, it is recommended to implement a beam-divergence control system to adapt the beam size dynamically to the unexpected movements of LAPS, achieving an optimal balance between fidelity and link availability. Alternatively, continuous-variable schemes using entangled states described in phase space may reduce background noise through coherent detection with a bright local oscillator, thereby deserving further investigations.
\subsection{LEO Satellite-to-Ground Quantum Links}
\begin{figure}[t]
\centering
\includegraphics[scale=0.58]{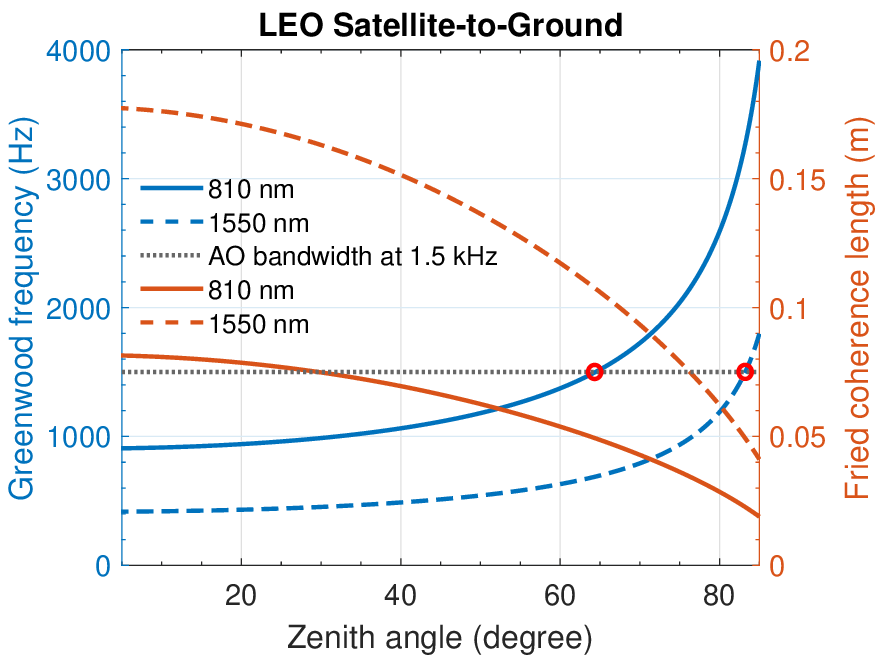}
\caption{Greenwood frequency and Fried coherence length versus zenith angle for a LEO satellite-to-ground downlink at 1550 nm and 810 nm.}
\label{fig_4}
\end{figure}
\subsubsection{Uplink/Downlink Challenges}
This scenario is indispensable to the integration of terrestrial and non-terrestrial quantum networks at intercontinental distances spanning thousands of kilometers. Consequently, it entails the longest atmospheric path, leading to increased turbulence effects in both uplink and downlink. Apparently, HAPS/LAPS-to-ground links also demonstrate similar attributes, albeit to a lesser extent, as they traverse shorter atmospheric paths. These links are typically employed to serve ground stations, covering distances ranging from several to hundreds of kilometers. Given that ground facilities are usually stationary, quantum links to the ground become increasingly vulnerable to outages due to intermittent cloud obstructions and extreme turbulence conditions. Another minor consideration involves the selection of optical wavelengths, such as 810 nm and 1550 nm, where 810 nm offers a balanced choice for atmospheric effects and cost-effective silicon-based SPADs, while 1550 nm, within the telecom band, aligns with standard fiber-based technology and facilitates energy-efficient miniaturized quantum transceivers using integrated silicon photonics. However, the choice of wavelength may vary depending on specific ground infrastructures and operational policies.
\subsubsection{Uplink/Downlink Solutions}
In order to overcome cloud blockage, it is essential to adopt a realistic model that captures the probability of a site being free from clouds at any given time \cite{birch2023}. This model should also consider additional outages caused by turbulence at all ground stations, enabling the application of site-diversity techniques to enhance connectivity with the most effective station. Ground stations commonly mitigate turbulence by employing large telescope apertures (0.4 m to 1.5 m), which diminish signal fluctuations through aperture-averaging effects achieved by spatially averaging the incoming light intensity over the receiver's aperture, thereby enhancing communication stability and reliability. The sophisticated superconducting nanowire single-photon detectors can also be deployed, necessitating a low-loss coupling of the free-space beam into a single-mode fiber (SMF). Achieving SMF coupling involves a fine-tracking subsystem integrated with adaptive optics (AO) \cite{gruneisen2021} for downlink beam stabilizations and corrections of aberrated wavefronts. This integrated subsystem is also utilized for pre-compensations of beam wander and phase distortions in the uplink, alleviating complex hardware on the satellites. The effectiveness of an AO system can be estimated by comparing the closed-loop control bandwidth with the Greenwood frequency that defines the temporal rate of change of turbulence. A desirable AO system should possess a closed-loop bandwidth matching the Greenwood frequency. Besides, turbulence intensity is gauged by the transverse spatial scale, known as the Fried coherence length. A longer Fried coherence length indicates milder turbulence with minimal wavefront distortions and vice versa.

In Fig. \ref{fig_4}, we provide a rigorous investigation of the Greenwood frequency \cite[(B1)]{gruneisen2021} and Fried coherence length \cite[(A2)]{gruneisen2021} using the corrected calculations of vertical turbulence and wind profiles with satellite slew rates presented in \cite{trinh2022} for a LEO-to-ground downlink. The satellite orbit is similar to Section \ref{subsect:LEO_HAPS_solutions}, and the ground station is located at the coordinates of the Institute of Industrial Science, The University of Tokyo, Japan. It is evident that the 1550-nm wavelength is more resilient against turbulence, visualized by a lower Greenwood frequency and a longer Fried coherence length with respect to that observed at 810 nm. The Greenwood frequency results also indicate that a state-of-the-art AO system with 1.5 kHz control bandwidth could effectively correct turbulence-induced wavefront aberrations at 1550 nm across all practical zenith angles below $80^{\circ}$ during the simulated satellite pass. In contrast, the benchmark AO system is only useful at zenith angles below $60^{\circ}$ when using the 810-nm wavelength.
%========================================%
\section{Future Directions}
\subsection{High-Dimensional Quantum Communications}
In quantum information processing, qubits are two-level quantum systems while \textit{qudits} are $d$-level systems that have $d$-dimensional state space, making them capable of storing and processing more information. Efficient manipulations of qudits could offer higher information capacity, enhanced security through multi-dimensional entanglements, and improved resilience to noise. More specifically, higher information capacity can be quantified using the relation $\log_2(d)$, which signifies the number of classical bits necessary to convey the same volume of information. For example, with $d=8$, 3 bits of classical information can be encoded using only one quantum state. When applied to entanglement-based QKD protocols, the transmission of multiple quantum states per photon significantly increases key generation rates, which is particularly beneficial for highly secure applications such as secured transactions in banking, public voting in government elections, and stealth operations in the military. Finally, qudits exhibit greater resilience to noise and decoherence due to redundant encoding of quantum information. This redundancy ensures that even if some states are affected, others remain intact, thus minimizing the overall impact of noise on the communication system. Recent progress has achieved quantum transport of high-dimensional spatial information with $d=15$ by using spatial modes of photons such as orbital angular momentum \cite{sephton2023}. Additional explorations into the generation, transmission, and detection of higher-dimensional qudits are necessary in both theoretical and experimental aspects to fully understand the fundamental limits of high-dimensional quantum states. Additionally, optimal designs and implementation of high-dimensional quantum terminals on NTN platforms are necessary to confirm the feasibility of this enhanced quantum feature towards the future Quantum Internet in the Sky.
\subsection{Multipartite Quantum Communications}
To realize a large-scale quantum network, simultaneous communications among multiple nodes are of great interest. In this context, multipartite quantum communications, involving the generation and distribution of entangled quantum states among multiple quantum systems, emerge as a viable solution that enriches dynamic configurations in the Quantum Internet. This is advantageous in quantum NTNs, where diverse connections across platforms are dynamically formulated and adjusted over time to accommodate on-demand ubiquitous quantum services. For instance, in distributed quantum computing, quantum terminals on various aerial platforms may share entangled states with distant ground facilities and perform quantum operations collectively to solve heavy computational tasks. In quantum secret sharing, a secret message is distributed among multiple parties in quantum NTNs in such a way that the secret message can only be reconstructed by a specific subset of selected parties. Combining multipartite features with high-dimensional quantum communications could enable the study of complex quantum correlations among multiple parties in higher-dimensional spaces, paving the way for advancements in distributed quantum computing, quantum cryptography, and quantum network architectures. Future research directions should, therefore, involve developing theoretical algorithms, experimental techniques, and network protocol architectures to harness the potential of multipartite high-dimensional quantum communications.
\subsection{Integrated Quantum Communications, Sensing, Computing, and Intelligence (IQCSCI)}
Practical applications of the future Quantum Internet would likely involve the integration of multifaceted quantum functionalities, fostering the integration of quantum communications, sensing, computing, and intelligence. In particular, quantum sensors embarked on non-terrestrial platforms leverage quantum principles to precisely detect environmental changes such as magnetic fields, gravitational forces, and acceleration at unprecedented accuracy levels. Additionally, quantum intelligence synergizes quantum computing and artificial intelligence to enhance the adaptability and decision-making capabilities of the Quantum Internet. One fascinating application of IQCSCI in quantum NTNs lies in empowering the autonomous driving of numerous vehicles on Earth. This is accomplished through the provision of secure communications, precise sensing, efficient computing, and intelligent decision-making capabilities. Quantum-secured communication links ensure data confidentiality, while quantum sensors enhance real-time environmental and traffic monitoring. High-dimensional quantum computing enables massive data processing, and quantum machine learning algorithms support adaptive decision-making. These advancements collectively optimize autonomous vehicle operation, navigation, and safety in diverse conditions. Future developments may delve further into performance trade-offs in IQCSCI functionalities, uncovering the underlying relationships and theoretical boundaries of this holistic vision.
%========================================%
\section{Conclusions}
In this article, we conceptualized the idea of a ``Quantum Internet in the Sky", envisioning a multi-layered network of aerial platforms, ranging from LEO satellites to HAPS and LAPS. This forms a sophisticated 3D mesh FSO communication infrastructure in the sky for ubiquitous quantum services. We then explored the distinctive challenges associated with these aerial platforms and delved into potential solutions, exemplified through detailed system designs and analyses. Lastly, we outlined crucial directions for future research in achieving a fully operational Quantum Internet, emphasizing the integration of high-dimensional multipartite quantum communications with sensing, computing, and intelligence.
\IEEEtriggeratref{7}
\bibliographystyle{IEEEtran}
\bibliography{IEEEabrv,References}
%%
%========================================%
\balance
\section*{Biographies}
\vspace{-33pt}
\begin{IEEEbiographynophoto}{Phuc V. Trinh}[M'17] received the B.E. degree in electronics and telecommunications from the Posts and Telecommunications Institute of Technology, Hanoi, Vietnam, in 2013, and the M.Sc. and Ph.D. degrees in computer science and engineering from The University of Aizu, Aizuwakamatsu, Japan, in 2015 and 2017, respectively. From 2017 to 2023, he was a Researcher with the Space Communication Systems Laboratory, National Institute of Information and Communications Technology, Tokyo, Japan. Since 2023, he has been a Project Research Associate with the Institute of Industrial Science, The University of Tokyo, Tokyo, Japan. His research interests are in optical and wireless communications for space, airborne, and terrestrial networks.
\end{IEEEbiographynophoto}

\begin{IEEEbiographynophoto}{Shinya Sugiura}[SM'12] received the B.S. and M.S. degrees in aeronautics and astronautics from Kyoto University, Kyoto, Japan, in 2002 and 2004, respectively, and the Ph.D. degree in electronics and electrical engineering from the University of Southampton, Southampton, U.K., in 2010. He was a Research Scientist with Toyota Central R\&D Laboratories, Inc., Japan, from 2004 to 2012 and an Associate Professor with Tokyo University of Agriculture and Technology, Japan, from 2013 to 2018. Since 2018, he has been an Associate Professor with the Institute of Industrial Science, The University of Tokyo, Tokyo, Japan. His research has covered wireless communications and signal processing. He authored or co-authored over 100 IEEE journal papers.
\end{IEEEbiographynophoto}

\vfill

\end{document}